\documentclass[cameraready]{Interspeech}

\title{Synthesizing the Lombard Effect: Multi-Level Control of Speech Clarity and Vocal Effort in TTS}

\author[affiliation={1,3}, orcid=0000-0003-1778-2136, correspondingauthor]{Seymanur}{Akti}
\author[affiliation={1,2}, orcid=0009-0001-0848-1000]{Alexander}{Waibel}

\address{
    $^1$ Karlsruhe Institute of Technology (KIT), Germany \\
    $^2$ Carnegie Mellon University (CMU), USA  \\
    $^3$ KIT Campus Transfer (KCT), Germany
}

\email{seymanur.akti@kit.edu, alexander.waibel@kit.edu}


\usepackage{comment}

\begin{document}

\maketitle

\begin{abstract}
    Humans tend to speak louder and clearer in challenging environments, such as noisy conditions or when addressing hearing-impaired listeners, which is called Lombard effect. To simulate this behavior in speech synthesis systems, we introduce a flow-matching based text-to-speech (TTS) model trained with vocal effort and articulation pseudo-labels. The proposed model achieves continuous and disentangled control of vocal effort and articulation, while also enabling word-level emphasis for clarifying specific segments of an utterance. Experimental results show that these control mechanisms effectively improve clarity-related acoustic features. Furthermore, speech-in-noise experiments demonstrate that our model successfully simulates the intelligibility gains of human clear speech in noisy conditions.
\end{abstract}

\section{Introduction}

Humans naturally adapt their speaking style depending on the listener and the acoustic environment. A prominent example is the Lombard effect, where speakers modify their speech in response to environmental noise or listener-related hearing difficulties. Lombard speech is typically characterized by increased vocal effort, higher pitch, slower speaking rate, flattened spectral tilt, and clearer articulation, all of which contribute to improved intelligibility under challenging listening conditions~\cite{junqua1993lombard, soltau2000specialized}. However, since these modifications deviate from neutral speech distributions, they challenge both speech processing systems~\cite{soltau2005compensating,soltau2002compensating} and the generative capabilities of text-to-speech applications. At the same time, simulating Lombard adaptations in text-to-speech (TTS) systems can improve intelligibility in noisy environments, support hearing-assistive technologies, and serve as a repair strategy in interactive spoken dialogue systems when communication failures occur~\cite{suhm1999model,constantin2022interactive}.

Despite the naturalness and effectiveness of these adaptations in human speech, most modern TTS systems are trained predominantly on non-Lombard speaking styles and therefore lack mechanisms for controllable Lombard-style speech production. As speech synthesis becomes increasingly integrated into real-world applications, including conversational agents and accessibility technologies, enabling interpretable and flexible control over Lombard characteristics is essential. Investigating how TTS models can incorporate these attributes represents an important step toward more robust, intelligible, and context-aware speech synthesis.

Early works on Lombard synthesis were mostly relied on signal-processing techniques such as spectral shaping~\cite{brouckxon2008time,sauert2012near,taal2013sii,taal2014speech}, dynamic range compression~\cite{zorila2014spectral,kandia2012speech}, manipulation of pitch, spectral tilt, and speaking rate \cite{raitio2011analysis}, and incremental synthesis frameworks for real-time transformation \cite{rottschafer2015online}. While effective, these approaches are limited in flexibility and often result in suboptimal naturalness.

More recent data-driven methods adapt neural TTS models to Lombard speech. Several works apply transfer learning on pretrained TTS systems using small Lombard datasets \cite{bollepalli2019lombard, paul2020enhancing}, or fine-tune models on whispered and Lombard speech \cite{hu2021whispered}. Other strategies include spectral tilt augmentation \cite{raitio2022vocal}, explicit pitch and energy modeling \cite{woszczyk2025voice}, ASR-guided loudness control \cite{novitasari2021dynamically}, and latent-space interpolation for continuous Lombardness control \cite{lobato2025gradual}.

Despite these advances, most prior work emphasizes global acoustic features such as SNR gain, vocal effort, or speaking rate. Synthesizing hyper-articulation, which is a key factor in intelligibility, remains rather underexplored. Earlier studies modeled articulation with HMM-based systems \cite{picart2020analysis}, while more recent work uses Global Style Tokens (GST) to capture hypo- and hyper-articulated speech patterns~\cite{nishihara2024low}. However, in existing work, articulation and vocal effort are typically handled independently, lacking a unified, disentangled control mechanism that captures their joint contribution. Motivated by this, we aim to design a system suitable for real-world scenarios, where acoustic conditions and listener needs vary dynamically, requiring flexible control over multiple speech attributes.

In this work, we propose a multi-dimensional Lombard TTS framework that jointly models vocal effort and articulation along separate control axes. Our main contributions are:
\begin{itemize}
\item A dual-axis conditioning framework that enables independent control of vocal effort and articulation, as well as their joint manipulation, allowing flexible synthesis.
\item A factorized injection strategy conditioning both the duration model and acoustic decoder, supporting temporal (speaking rate, phoneme duration) and spectral–prosodic (energy, spectral tilt) control.
\item Multi-level control, supporting both utterance- and word-level adjustments for targeted intelligibility enhancement.
\end{itemize}

We share the sample audios at the demo page~\footnote{\href{https://seymanurakti.github.io/synthesizing-lombard-effect/}{https://seymanurakti.github.io/synthesizing-lombard-effect/}}

\section{Method}

\subsection{Model Architecture}

We adopt Matcha-TTS~\cite{mehta2024matcha} as our base architecture due to its strong balance between synthesis quality and inference efficiency. Matcha-TTS is a flow-matching-based TTS model that replaces iterative diffusion sampling with a deterministic flow prediction objective, significantly reducing inference time while maintaining competitive audio quality.

The decoder is an Optimal Transport (OT)-based flow matching decoder. The continuous nature of flow matching supports smooth interpolation in acoustic space, aligning with our goal of gradual and multi-dimensional control over vocal effort and articulation. Matcha-TTS further incorporates Monotonic Alignment Search (MAS) to learn phoneme-to-frame durations, allowing explicit control over duration-related features. Since Lombard speech involves temporal adaptations such as reduced speaking rate and modified articulation patterns, phoneme-level duration modeling provides a controllable link between text and acoustics.

\begin{figure}
    \centering
    \includegraphics[width=\linewidth]{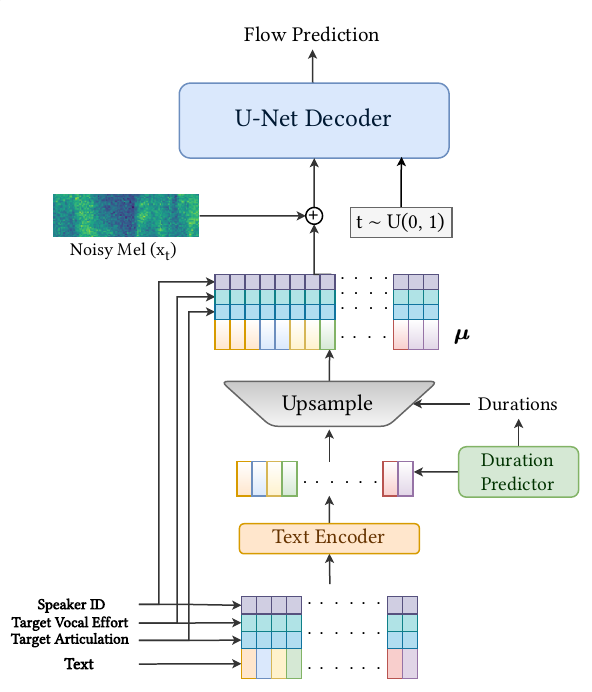}
    \caption{Overall architecture of Matcha-TTS with multi-level vocal effort and articulation conditioning.}
    \label{fig:arch}
    \vskip -0.5cm
\end{figure}

We follow the original training recipe of Matcha-TTS. As for vocoder, we use Vocos~\cite{siuzdak2023vocos} for high-fidelity and efficient waveform reconstruction. The overall training pipeline is illustrated in Figure~\ref{fig:arch}.

\subsection{Feature Conditioning and Multi-Level Control}

To enable reference-free and flexible style control, we introduce dual conditioning that jointly models articulation and vocal effort. Using the predefined style classes of the Expresso~\cite{nguyen2023expresso} dataset, default (neutral), enunciated (hyper-articulated), fast (hypo-articulated), and projected (increased vocal effort), we derive pseudo-labels for the articulation and vocal effort dimensions. Articulation is modeled along the fast–enunciated axis, while vocal effort is modeled along the default–projected axis.
Each discrete label is mapped into a 32-dimensional continuous embedding space via learnable linear projection layers. The resulting articulation and vocal effort embeddings are concatenated together as a joint embedding, enabling both independent control of each attribute and their joint manipulation. To incorporate multiple speakers, speaker identity is represented using discrete embeddings of the same dimensionality and concatenated with the style embeddings.

As both articulation and vocal effort influence temporal and spectral properties of speech, we adopt a dual-injection strategy that conditions both duration modeling and acoustic generation:
\begin{itemize}
    \item \textbf{Encoder-side injection (duration control):} Style and speaker embeddings are concatenated to the text encoder inputs before encoding and duration prediction, enabling control over speaking rate, phoneme stretching, and articulation patterns.
    \item \textbf{Decoder-side injection (acoustic control):} The same embeddings are concatenated to the U-Net-based flow-matching decoder inputs, allowing modulation of spectral tilt, energy distribution, and formant clarity while preserving speaker information throughout generation.
\end{itemize}

Embeddings are temporally broadcast by repeating them across text tokens for the encoder and mel frames for the decoder. This token-level conditioning naturally enables fine-grained control of articulation during inference in addition to the word-level emphasis annotations from the Expresso dataset, allowing localized clarity enhancement for selected segments of an utterance.

\subsection{Inference-time Control}

During inference, vocal effort ($\alpha \in [0, 1]$) and articulation ($\beta \in [0, 1]$) are treated as continuous controllable scalars. By interpolating within the learned embedding spaces, the model can synthesize speech with gradually varying clarity, simulating adaptive responses under diverse acoustic conditions. This enables independent adjustment of articulation and vocal effort, smooth transitions between clarity levels, and fine-grained adaptation to listener or noise requirements.

For the word level control, different $\beta$ values can be assigned across the time axis. Higher $\beta$ values can be applied to target words for increased articulation, while emphasis labels present in the training data enhance their acoustic prominence.

\begin{figure*}[!]
    \centering
    \includegraphics[width=\textwidth]{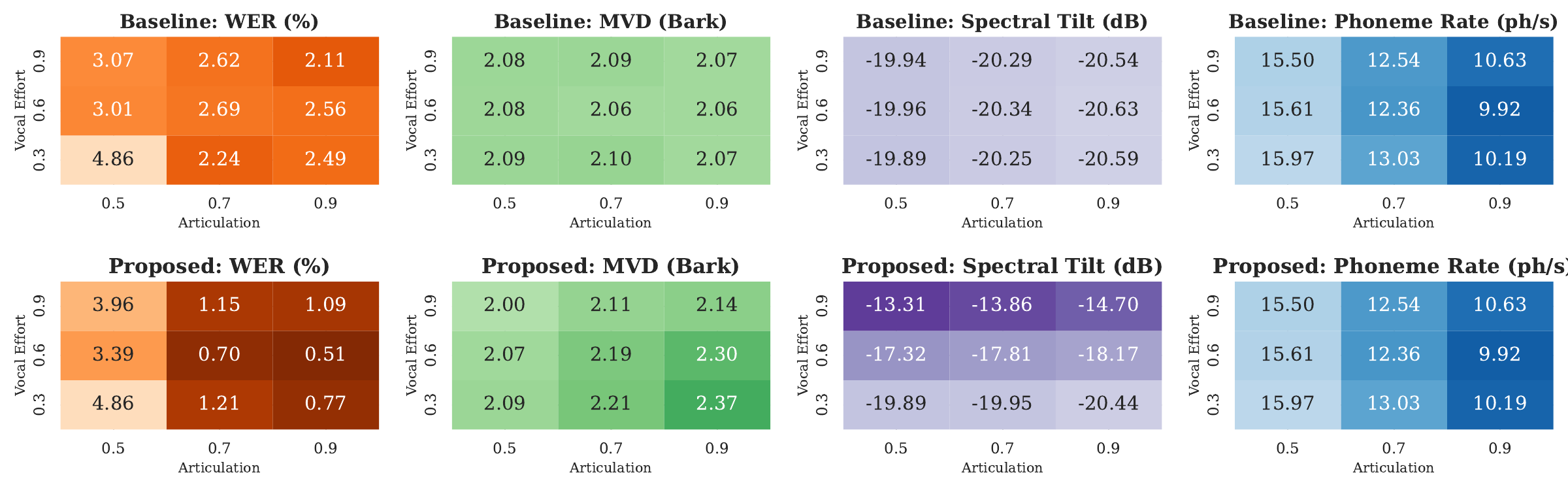}
    \caption{Comparison to the baseline in terms of WER $\downarrow$, MVD $\uparrow$, spectral tilt $\uparrow$ and phoneme rate $\downarrow$.}
    \label{fig:baseline_comparison}
    \vskip -0.3cm
\end{figure*}

\section{Experiments and Results}

\subsection{Training Data}
We train our model using a subset of the Expresso dataset containing the default, enunciated, fast, and projected speaking styles, resulting in approximately 11 hours of speech including  four speakers (two female and two male).

Since the total duration of the Expresso subset is relatively limited for learning comprehensive phoneme coverage and robust pronunciation modeling, we augment the training data with LJ Speech. The LJ Speech corpus is treated as a neutral speaker with mid-level articulation and baseline vocal effort. Incorporating this dataset improves phonetic diversity and stabilizes training, while preserving the controllable style structure learned from Expresso. We used pseudo labels as following:
\begin{itemize}
    \item Articulation Labels: 0.1 (fast) 0.5 (neutral) 0.9 (enunciated)
    \item Vocal Effort Labels: 0.3 (neutral) 0.9 (projected)
\end{itemize}
\subsection{Evaluation Metrics}

We evaluate our system using the Harvard Sentences dataset~\cite{rothauser1969ieee}, which provides phonetically balanced sentences suitable for articulation and intelligibility analysis. The first five lists (50 sentences per speaker) were synthesized under varying articulation ($\beta$) and vocal effort ($\alpha$) levels.

As a comparison, we construct a naive signal-processing baseline. Vocal effort is simulated by RMS matching neutral speech to higher-effort samples, while decreased speaking rate is approximated using linear time-stretching. This baseline allows us to isolate the benefit of learned style control over simple amplitude and rate manipulation.

We report the following objective metrics:

\begin{itemize}
    \item \textbf{Word Error Rate (WER):} Computed using Whisper-medium\footnote{https://huggingface.co/openai/whisper-medium} to assess intelligibility of synthesized speech.
    \item \textbf{Spectral Tilt:} Defined as the ratio of spectral energy between 5-1kHz to energy below 1kHz~\cite{murphy2008investigation}. Increased vocal effort is expected to shift energy toward higher frequencies resulting in higher spectral tilt.
    \item \textbf{Mean Vowel Dispersion (MVD):}  as the Euclidean distance of /i/, /a/, and /u/ vowel formants to the vowel centroid in F1-F2 space. Higher dispersion indicates clearer articulation.
    \item \textbf{Phoneme Rate:} Measured as phonemes per second. Hyper-articulated speech is expected to reduce speaking rate.
    \item \textbf{Speech Intelligibility Index (SII):} Defines the percentage of speech information audible to a listener given the noise.
\end{itemize}

\subsubsection{Human Evaluation}

We conducted a comparative Mean Opinion Score (CMOS) study with 10 participants. Participants were given 9 questions for naturalness and 18 questions for intelligibility.

\textbf{Naturalness:} Participants compared the baseline with Lombard-style synthesized speech and rated their preference on a $[-3, +3]$ scale, where $-3$ indicates strong preference for the baseline and $+3$ indicates strong preference for our method.

\textbf{Intelligibility:} Participants compared neutral speech with Lombard-style speech under noisy conditions using the same $[-3, +3]$ scale, where positive values indicate higher intelligibility of Lombard speech. Participants were not informed which sample corresponded to which condition.

\subsection{Results and Discussion}

\subsubsection{Analysis of Lombard-Related Signal Changes}

For the analysis of the Lombard effect, we first report the signal-level changes achieved by our model and compare them with the baseline. Figure~\autoref{fig:baseline_comparison} shows the variations in WER, Spectral Tilt, Mean Vowel Dispersion (MVD), and Phoneme Rate across different Vocal Effort ($\alpha$) and Articulation ($\beta$) levels.

We observe that WER decreases consistently as $\beta$ increases, indicating that articulation has a strong impact on intelligibility under default conditions. In contrast, the baseline system shows marginal WER improvements, suggesting that linearly slowing down speech alone is suboptimal for intelligibility enhancement. Along the $\alpha$ axis, we also observe improvements, however, at $\alpha = 0.9$, performance drops slightly, likely due to ASR sensitivity to highly projected speech, which deviates from its training distribution also shown in~\cite{tuttosi2026bersting}. 

Mean Vowel Dispersion increases with $\beta$, reflecting more distinct vowel articulation in hyper-articulated speech. Across the $\alpha$ axis, MVD remains relatively stable, indicating that vocal effort changes do not significantly affect articulation patterns. 

For Spectral Tilt, the largest variation is observed along the $\alpha$ axis. Increasing vocal effort shifts energy toward higher frequencies compared to the baseline, where only RMS gain is applied and energy is amplified uniformly across frequencies. Along the $\beta$ axis, Spectral Tilt remains largely consistent, supporting independent control.

\subsubsection{Speech in Noise Experiments}

We evaluate intelligibility under clean and noisy conditions (SNR = 10, 5, 1) using three noise types: restaurant babble, overlapping speech, and white noise to analyze how articulation and vocal effort interact with different masking characteristics. During mixing, noise levels are adjusted according to speech RMS with fixed SNR across lombardness levels to ensure differences arise from speaking style rather than simple SNR gain due to amplification.

\begin{figure}[h]
    \centering
    \includegraphics[width=\linewidth]{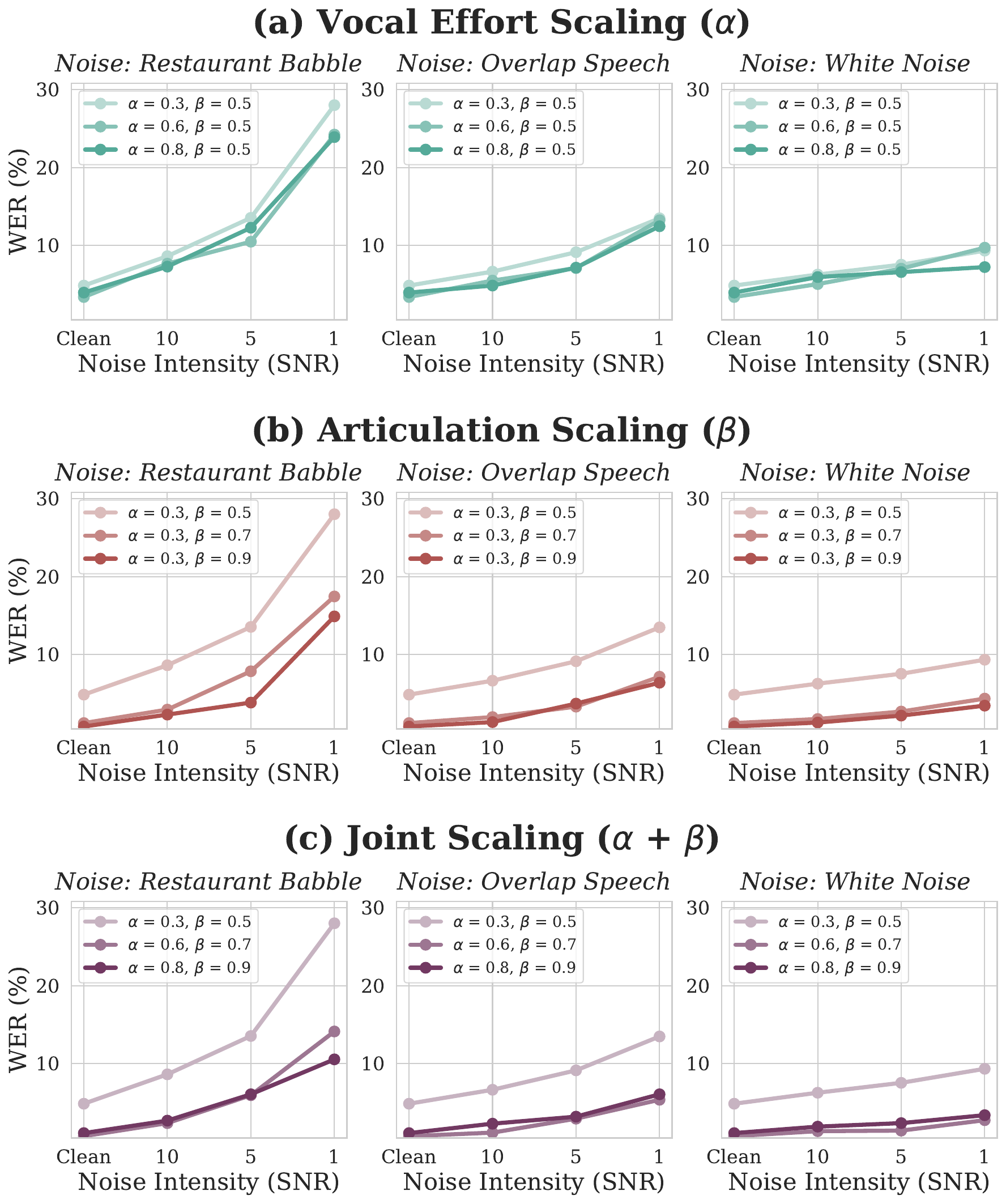}
    \caption{WER of synthesized speech under noisy conditions.}
    \label{fig:wer_line_plots}
\end{figure}

As shown in Figure~\ref{fig:wer_line_plots}(b), increasing articulation ($\beta$) consistently reduces WER across noise types, with strongest gains observed in mid-level noises. Improvements saturate beyond mid-to-high articulation levels. Vocal effort ($\alpha$) provides minimal WER gains under RMS-normalized evaluation, since its contribution arises primarily from spectral redistribution rather than amplitude increase. Nevertheless, it still contributes to intelligibility under higher noise levels, particularly for restaurant babble and white noise.

\begin{figure}[h]
    \centering
    \includegraphics[width=\linewidth]{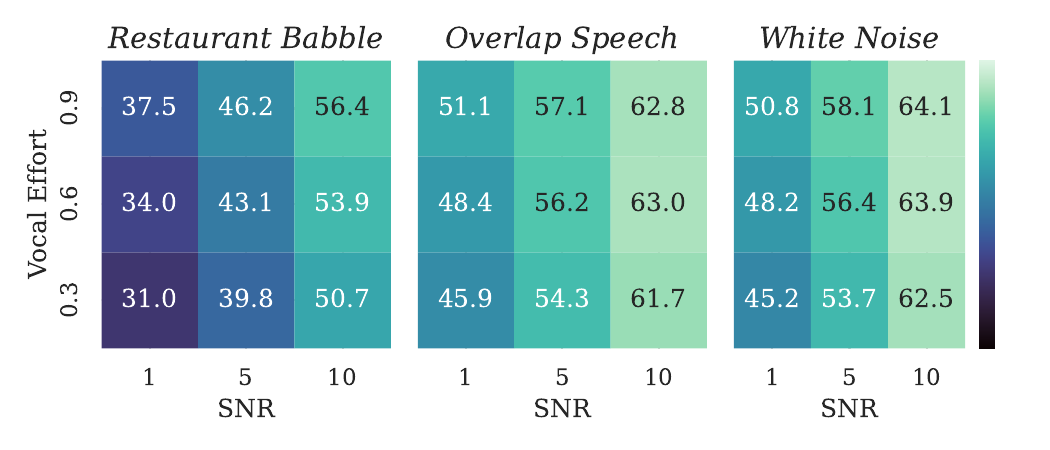}
    \caption{Average SII (\%) results. Higher the better.}
    \label{fig:sii_results}
\end{figure}

For WER analysis, we restrict $\alpha$ to 0.8, as extreme projection ($\alpha = 0.9$) introduces ASR degradation due to distribution mismatch, harming the interpretable trends. Since WER may under-represent the benefits of vocal effort, we additionally report SII as an audibility-based metric. As shown in Figure~\ref{fig:sii_results}, increased vocal effort improves audibility across noise types, with the largest gains in restaurant babble at low SNR levels and slightly smaller gains in white noise and overlapping speech.

Joint scaling of articulation and vocal effort provides the largest gains under severe noise (SNR = 1), particularly in restaurant babble. While mid-level articulation alone yields substantial improvements in white noise and overlapping speech under fixed-SNR conditions, combining articulation with vocal effort offers additional robustness in highly masked scenarios. These results indicate complementary contributions: articulation enhances phonetic distinctiveness across conditions, whereas vocal effort improves spectral audibility under energetic masking.

\subsubsection{Human Evaluation Results}

We report the subjective evaluation results from 10 participants in the Table~\ref{tab:cmos} highlighting the perceptible intelligibility gain in noisy speech and naturalness superiority over baseline. Detailed analysis shows that the participants consistently rated the baseline time-stretched speech as unnatural, indicating that simple rate manipulation is insufficient for natural Lombard synthesis. For noise types, vocal effort has shown the best effect on restaurant babble and articulation consistently improve intelligibility across noise levels.

\begin{table}[h]
\small
\centering
\caption{CMOS in [-3, +3] with 95\% confidence intervals.}
\label{tab:cmos}
\begin{tabular}{lc}
\toprule
\textbf{Task} & \textbf{CMOS $\uparrow$} \\
\midrule
Naturalness       & 1.97 $\pm$ 0.32\\
Intelligibility        & 1.13 $\pm$ 0.24\\
\bottomrule
\end{tabular}
\vskip -0.4cm
\end{table}

\subsubsection{Word-Level Emphasis}

To emphasize a target word, we assign a higher articulation index ($\beta$) to its tokens while reducing $\beta$ for surrounding tokens as a sufficiently large contrast in articulation is required for perceptual salience. We therefore assign $\beta = 1.5$ to the emphasized word and $\beta = 0.1$ to neighboring tokens. This slows the target word while maintaining near-baseline phoneme rate for surrounding words. Additionally, emphasis labels are used to add more acoustic stress on the words.

We maintain a sufficiently large articulation index difference for the targeted word for the effect to be perceptible. Thus, we create a peak at the desired timestep where emphasized word itself is assigned $\beta = 1.5$. To counteract articulation leakage into neighboring tokens, we assign a low $\beta = 0.1$ to surrounding words. This preserves the baseline phoneme rate (15.8 ph/s) while only slowing the emphasized word (10.5 ph/s), ensuring localized temporal modulation.

For evaluation, we use ASR predictions on neutral baseline speech ($\beta = 0.5$, $\alpha = 0.3$) to identify transcription errors. Only samples containing incorrectly recognized words are resynthesized with increased articulation. We report WER for these samples in Table~\ref{tab:word_level_wer} where results indicate that hyper-articulation improves intelligibility more than emphasis alone, while combining both yields the largest reduction in WER.

\begin{table}[h]
\centering
\small
\caption{WER on previously mistaken words after word-level emphasis and hyper-articulation.}
\label{tab:word_level_wer}
\begin{tabular}{lc}
\toprule
\textbf{Condition} & \textbf{WER (\%) $\downarrow$} \\
\midrule
Baseline (neutral)       & 17.61 \\
\midrule
Hyper-articulation only        & 6.81  \\
Emphasis only            & 9.15  \\
Both combined            & \textbf{3.90}  \\
\bottomrule
\end{tabular}
\vskip -0.4cm
\end{table}

\section{Conclusion}

We proposed a multi-dimensional Lombard speech synthesis approach that jointly models vocal effort and articulation in a controllable TTS framework. Using flow-matching-based TTS and dual-level conditioning, the system enables continuous control over global speaking style and word-level emphasis, yielding complementary gains in intelligibility, particularly under complex noise such as restaurant babble. Evaluations show articulation primarily drives intelligibility, while vocal effort enhances spectral clarity and audibility. Word-level emphasis further improves intelligibility for targeted segments. Overall, our model can effectively simulate human-like Lombard strategies, providing flexible tools for context-aware speech synthesis.

\textbf{Limitations and Future Work}—ASR-based evaluation may underestimate the benefits of Lombard speech due to sensitivity to extreme vocal effort. Additionally, token-level control can be  refined to reduce leakage between words. Future work will explore more robust evaluation metrics, improved fine-grained control, and real-time adaptation to dynamically changing acoustic conditions.

\section{Acknowledgments}
This research is supported by the European Union’s Horizon Europe programme grant agreement No. 101213369 (DVPS) and KIT Campus Transfer GmbH (KCT) in accordance with the collaboration with Carnegie-AI.

\section{Generative AI Use Disclosure}
During the preparation of this manuscript, the authors utilized generative AI technologies solely for the purposes of language polishing, grammar verification, and structural formatting of the content.

\bibliographystyle{IEEEtran}
\bibliography{mybib}

\end{document}